# Improved El Niño-Forecasting by Cooperativity Detection

Josef Ludescher[1], Avi Gozolchiani[2], Mikhail I. Bogachev[1,3], Armin Bunde[1], Shlomo Havlin[2], and Hans Joachim Schellnhuber[4,5]

[1]Institut für Theoretische Physik, Justus-Liebig-Universität Giessen, 35392 Giessen, Germany
[2]Minerva Center and Department of Physics, Bar-Illan University, 52900, Ramat Gan, Israel
[3]Radio Systems Department, St. Petersburg Electrotechnical University, 197376, St. Petersburg, Russia
[4]Potsdam Institute for Climate Impact Research, 14412 Potsdam, Germany
[5]Santa Fe Institute, 1399 Hyde Park Rd. Santa Fe, NM 87501, USA

**Abstract**

Although anomalous episodical warming of the eastern equatorial Pacific, dubbed El Niño by Peruvian fishermen, has major (and occasionally devastating) impacts around the globe, robust forecasting is still limited to about six months ahead. A significant extension of the pre-warming time would be instrumental for avoiding some of the worst damages such as harvest failures in developing countries. Here we introduce a novel avenue towards El Niño-prediction based on network methods inspecting emerging teleconnections. Our approach starts from the evidence that a large-scale cooperative mode - linking the El Niño-basin (equatorial Pacific corridor) and the rest of the ocean - builds up in the calendar year before the warming event. On this basis, we can develop an efficient 12 months-forecasting scheme, i.e., achieve some doubling of the early-warning period. Our method is based on high-quality observational data as available since 1950 and yields hit rates above 0.5, while false-alarm rates are below 0.1.

The so-called El Niño-Southern Oscillation (ENSO) is the most important phenomenon of contemporary natural climate variability [1-4]. It can be perceived as a self-organized dynamical see-saw pattern in the Pacific ocean-atmosphere system, featured by rather irregular warm (El Niño) and cold (La Niña) excursions from the long-term mean state. ENSO has considerable influence on the climatic and environmental conditions in its core region, but affects also distant parts of the world. The pattern is causing disastrous floodings in countries like Peru and Ecuador as well as heavy droughts in large areas of South America, Indonesia and Australia. It is arguably also associated with severe winters in Europe, anomalous monsoon dynamics in East Asia, intensity of tropical cyclones such as hurricanes in the Carribean, and epidemic diseases occurring in a variety of places [5-9].

Strong El Niño-events, in particular, have affected, time and again, the fate of entire societies. A popular, yet quite informative account of ENSO's destructive power is provided in Ref. 10. This book investigates the pertinent droughts in India, China and Brazil towards the end of the 19[th] century, which killed an estimated 30-50 million people.

What happened in pre-modern times is unlikely to be repeated in the future. However, anthropogenic global warming [11-12] may have a significant effect on the character of ENSO and render this geophysical pattern even more challenging for certain societies. In fact, the phenomenon is listed among the so-called "tipping elements" in the Earth System [13-14] that might be transformed – sooner or later – by the greenhouse-gas emissions from fossil-fuel burning and land-cover change. The scientific jury is still out, pondering the question how El Niño-events will behave in a world without aggressive climate-protection measures [15]. Will the eastern tropical Pacific warm permanently, periodically, or as irregularly as nowadays? Will the oscillation go away completely (something that appears rather unlikely according to Ref. 4) or gain in strength (as suggested by some paleo-climatic data)? In the latter case, anything that helps to improve the predictive power of the scientific ENSO analysis would be even more important than it is already today.

The ENSO phenomenon is currently tracked and quantified, for example, by the NINO3.4 index, which is defined as the average of the sea-surface temperature (SST) anomalies at certain grid points in the Pacific (see Fig. 1). An El Niño-episode is said to occur when the index is above 0.5°C for a period of at least 5 months. Sophisticated global climate models taking into account the atmosphere-ocean coupling as well as dynamical systems approach, autoregressive models and pattern-recognition techniques applied on observational and reconstructed records have been used to forecast the pertinent index with lead times between 1 and 24 months. Up to 6 months, the various forecasts perform reasonably well, while for longer lead times the performance becomes rather low [16-29]. A particular difficulty for prediction of the NINO3.4 index is the «spring barrier» (see, e.g., [30]). During boreal spring time, anomalies that develop randomly in the western Pacific, reduce the signal-to-noise ratio for the dynamics relevant to ENSO, and make it harder to predict across the barrier.

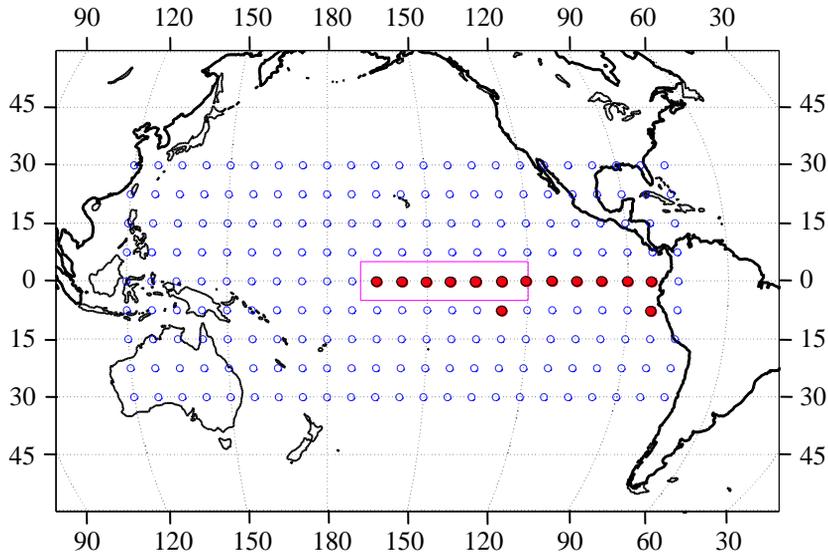

Figure 1: The "climate network". Each node inside the El Niño basin (full red symbols) is linked to each node outside the basin (open symbols). The nodes are characterized by their air temperature at the surface level (SAT), and the link strength between the nodes are determined from their cross-covariance. The red rectangle denotes the area where the NINO3.4 index is measured. For the definition of the El Nino basin [32], we have followed [33,35]. In the supplemental material, we provide a sensitivity test for this choice and show, for example, that the inclusion of the two nodes south of the Equator is not essential for our results.

In this study, we follow a different route. Instead of considering the time dependence of climate records at single grid points $i$, we study the time evolution of the interactions (teleconnections) between pairs of grid points $i$ and $j$, which are represented by the strengths of the cross correlations between the climate records at these sites. The interactions can be considered as links in a climate network where the nodes are the grid points [31,33,35]. Recent empirical studies have shown that in the large-scale climate network the links tend to *weaken* significantly *during* El Niño-episodes, and this phenomenon is most pronounced for those links that connect the "El Niño-basin" (full circles in Fig. 1) [32] with the surrounding sites in the Pacific ocean (open symbols) [33,35].

Therefore we concentrate on these links and show that well *before* an El Niño-episode their mean strength tends to *increase*. We use this robust observation to forecast El Niño-development more than 1 year in advance. We utilize the time span between 1950 and 2011, where the information on ENSO dynamics is reliable, and the observational data necessary for constructing the climate network is complete [36].

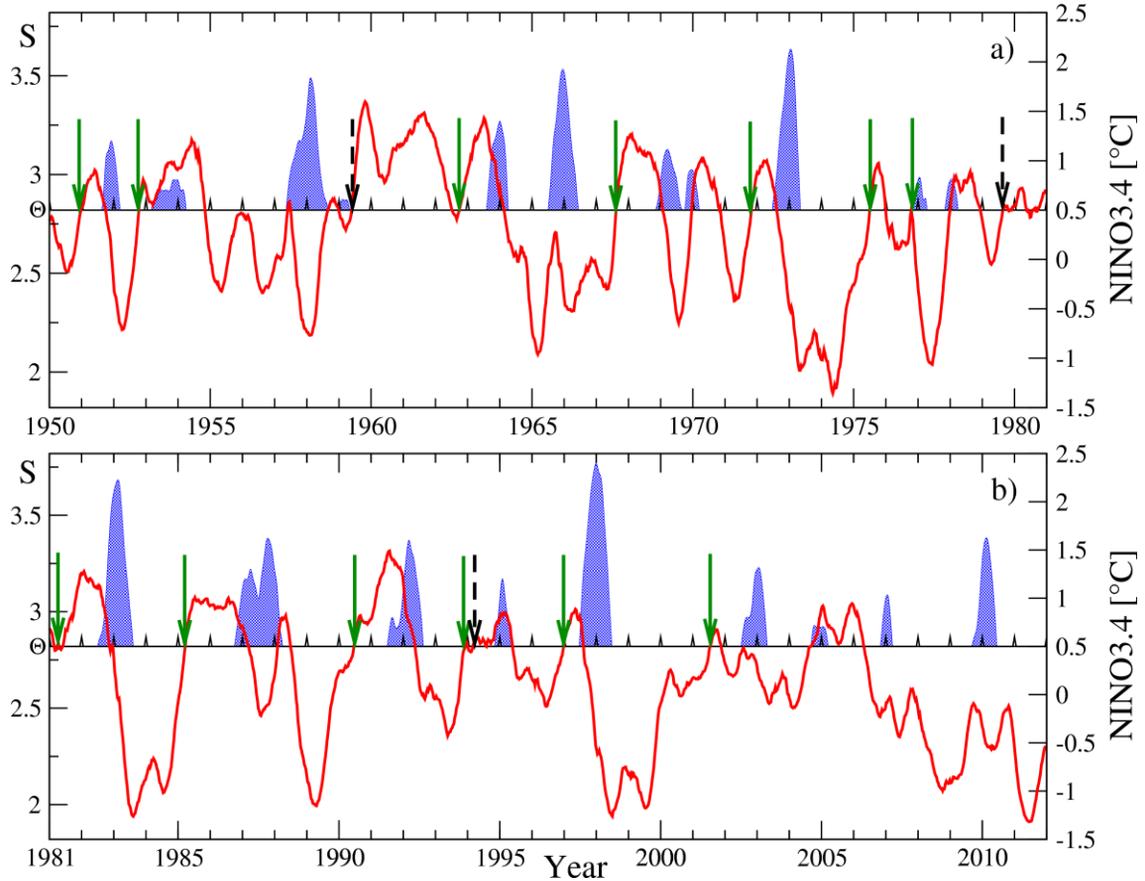

Figure 2: **The forecasting algorithm.** We compare the average link strength $S(t)$ in the climate network (red curve) with a decision threshold $\Theta$ (horizontal line, here $\Theta = 2.82$) (left scale) with the standard NINO3.4 index (right scale), between January 1, 1950 and December 31, 2011. When the link strength crosses the threshold from below outside an El Niño episode, we give an alarm and predict that an El Niño episode will start in the following calendar year. The El Niño episodes (when the NINO3.4 index is above 0.5°C for at least 5 months) are shown by the filled blue areas. The first half of the record (a) is the learning phase where we optimize the decision threshold. In the second half (b), we use the threshold obtained in (a) to predict El Niño episodes. Correct predictions are marked by green arrows and false alarms by dashed arrows. In order to resolve by eye the accurate positions of the alarms, we show in the supplemental material (Fig. 5) magnifications of those parts of Fig. 2 where the crossings or non-crossings are difficult to see clearly without magnification. We also show the alarms for the slightly smaller threshold $\Theta = 2.81$ (Fig. 6 in the supplemental material), which yields the same performance in the learning phase and one more false alarm in the prediction phase. The lead time between the prediction and the beginning of the El Nino episodes is 0.94±0.44y, while the lead time to the maximal NINO3.4 value is 1.4±0.33y.

We employ the network shown in Fig. 1, which consists of 14 grid points in the El Niño-basin and 193 grid points outside this domain. Following [35], we consider at each node $k$ the daily atmospheric temperature anomalies (actual temperature value minus climatological average for each calendar day) at the surface area level, and define the relative temperature anomaly $\theta_k(t) = \left(T_k(t) - \langle T_k(t) \rangle\right) / \left\langle \left(T_k(t) - \langle T_k(t) \rangle\right)^2 \right\rangle^{1/2}$, where the brackets denote an average over the past 365 days, according to $\langle f(t) \rangle = \frac{1}{365} \sum_{m=0}^{364} f(t-m)$. The data have been obtained from the NCEP/NCAR Reanalysis I project [36, 37].

For obtaining the time evolution of the strengths of the links between the nodes $i$ inside the El Niño-basin and the nodes $j$ outside we compute, for each 10*th* day $t$ in the considered time span between 1950 and 2011, the time-delayed cross covariance function defined as $C_{ij}^{(t)}(-\tau) \equiv \langle \theta_i(t)\theta_j(t-\tau) \rangle$ and $C_{ij}^{(t)}(\tau) \equiv \langle \theta_i(t-\tau)\theta_j(t) \rangle$. We consider time lags τ between 0 and 200d, where a reliable estimate of the background noise level can be guaranteed. Note that for estimating the cross-covariance function at day *t*, only temperature data from the past are considered. A representative example of $C_{ij}^{(t)}(\tau)$ is shown in the supplemental material (Fig. 4).

Next we determine, for each point in time *t*, the maximum, the mean, and the standard deviation of $C_{ij}^{(t)}(\tau)$ around the mean and define the link strength $S_{ij}(t)$ as the difference between maximum and mean value, divided by the standard deviation. Accordingly, $S_{ij}(t)$ describes the link strength at day *t* relative to the underlying background, and thus quantifies the dynamical teleconnections between nodes *i* and *j*. We obtain the desired mean strength *S* (*t*) of the dynamical teleconnections in the climate network by simply averaging over all individual link strengths. In this average, we do not weight nodes from different latitudes according to their density, since the range of such weights varies insignificantly for the narrow range of latitudes depicted in our network.

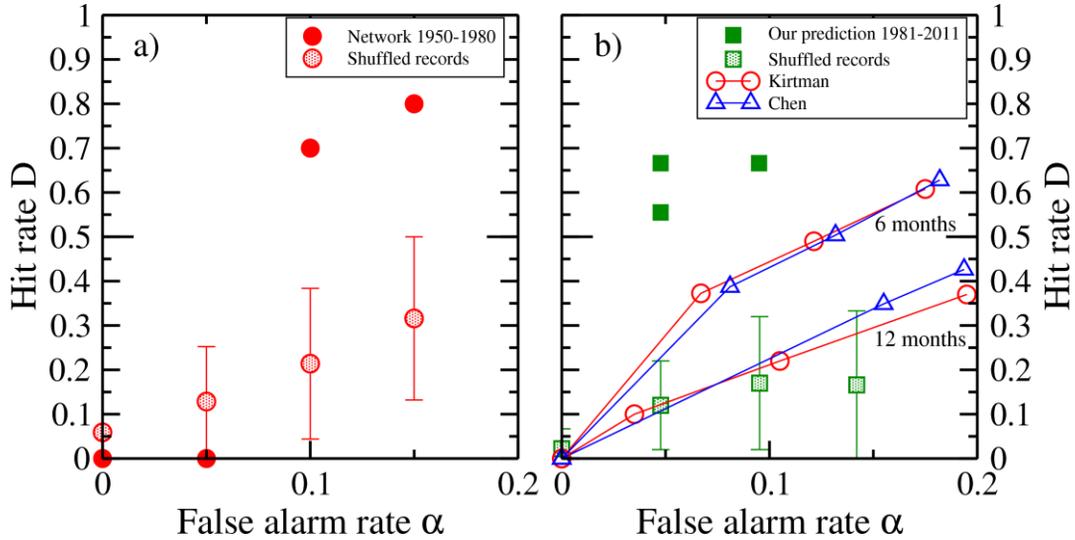

Figure 3: **The prediction accuracy.** (a) shows for the four lowest false alarm rates α = 0, 0.05, 0.1 and 0.15, the best hit rates *D* in the learning phase (see Fig. 2a). The best results are obtained at α = 0.1 and 0.15. For α = 0.1, the decision threshold Θ is between 2.805 and 2.822. For α = 0.15, Θ is between 2.780 and 2.792. The results for the randomized *S*(*t*) with error bars are shown by the shaded circles. (b) shows the quality of the prediction in the second half of the record, when the above thresholds are applied. For 2.816 < Θ ≤ 2.822, we have *D* = 0.667 at α = 0.048, for 2.805 < Θ ≤ 2.816, we have *D* = 0.667 at α = 0.095, and for 2.780 < Θ ≤ 2.792, we have *D* = 0.556 at α = 0.048. For comparison, we show also results for 6- and 12-months forecasts based on climate models [21, 38]. The shaded squares and the error bars denote the mean hit rates and their standard deviations for predictions based on the shuffled data.

Figure 2 depicts the time evolution of *S*(*t*) between January 1, 1950 and December 31, 2011. To statistically validate our method, we have divided this time interval into two equal parts. The first part (up to 31 December 1980, upper panel) is used for learning the optimum prediction algorithm. In the second part (from January 1, 1981 up to December 31, 2011, lower panel) we apply this algorithm to predict the El Niño-episodes. The figure compares the time dependence of *S*(*t*) (left scale) with the standard NINO3.4 index (right scale). The El Niño-episodes where the index is above *C* for at least 5 months are marked in blue. The figure shows that during an El Niño-event, the mean strength *S* of the interactions tends to decrease, supporting the hypothesis that the El Niño-basin tends to decouple from the rest of the globe when the anomalous warming is in full swing [35].

More relevant, from the perspective of forecasting, is the finding that well before an episode $S(t)$ tends to increase, i.e., the *cooperativity* between the El Niño-basin and the surrounding sites in the Pacific area grows. This feature is used here for predicting the start of an El Niño-event in the following year. To this end, we place a varying horizontal threshold $S(t) = \Theta$ in the upper panel of Fig. 2 and mark an alarm when $S(t)$ crosses the threshold from below *outside* an El Niño-episode (i.e., when the NINO3.4 index is below $0.5°C$). We assume that such an alarm forecasts an El Niño to develop in the following calendar year. If there are multiple alarms in the same calendar year, only the first one is regarded. The alarm results in a correct prediction, if in the following calendar year an El Niño-episode actually sets in; otherwise it is regarded as a false alarm.

For illustrating the algorithm, we shifted the $S(t)$ curve in Fig. 2 vertically such that the El Niño-threshold ($0.5°C$) coincides with our chosen threshold (here $\Theta = 2.82$). Correct predictions are marked by green arrows and false alarms by dashed arrows. Between 1951 and 1980, there are 10 years where an El Niño-episode started (i.e., there are 10 events), and 20 years where it did not start (i.e., there are 20 non-events). In the figure, we see 7 correct predictions and 2 false alarms, giving rise to the hit rate 7/10 and the false-alarm rate 2/20, respectively. By altering the magnitude of the threshold, we vary the hit rate and the false-alarm rate. Figure 3a shows, again for the learning period between 1950 and 1980, the best hit rates for the (tolerable) false alarm rates 0, 1/20, 2/20, and 3/20. The best performances are for thresholds $\Theta$ in the interval between 2.805 and 2.822, where the false-alarm rate is 2/20 and the hit rate is 0.7, and for thresholds between 2.780 and 2.792, where the false-alarm rate is 3/20 and the hit rate is 0.8.

For demonstrating that these results are not accidental, we analyzed randomized $S(t)$ curves obtained by reshuffling the temperature records at each site. We only randomized the calendar years but not the data within each calendar year. In this way we preserved the short-term memory in each record but reduced the cross-correlations between them. We considered 100 such randomizations and determined for each of them, for the false-alarm rates 0, 1/20, 2/20, and 3/20, respectively, the best hit rates. To characterize the distribution of the best hit rates, we calculated their mean and standard deviation. The results, also shown in Fig. 3a, are well below the hit rates achieved with the observational $S(t)$ curve.

Next, we use the thresholds selected in the learning phase to predict El Niño-episodes in the second half of the data set between 1982 and 2011, where we have 9 episodes and 21 non-episodes. For $\Theta$ between 2.816, and 2.822, which is depicted in Fig. 2b, the hit rate is $D = 6/9 \cong 0.667$ – at a false-alarm rate $\alpha = 1/21 \cong 0.048$. For $\Theta$ between 2.805 and 2.816, the hit rate is also $D = 6/9$, but the false-alarm rate is $\alpha = 2/21 \cong 0.095$. For $2.780 < \Theta \leq 2.792$, we have $D = 5/9 \cong 0.556$ at $\alpha = 1/21 \cong 0.048$. These results are highly significant since the prediction efficiency is considerably better than for the shuffled data.

For comparison, we show also the results for 6- and 12-months forecasts based on state-of-the-art climate models [21, 38]. In [21], an ensemble of model trajectories have been used, while for the forecast of [38], only a single trajectory has been employed. In both references, the forecast has been compared with the NINO3.4 index, as in the current analysis. Figure 3b shows that the method suggested here for predicting El Niño-episodes more than 1 year ahead considerably outperforms the conventional 6-months and 1-year forecasts. It should be noted that while one can tune lead time and robustness in physical models, this is not possible in our statistical predictions. In contrast to physical models, which predict the SST values in the relevant regions and use them for a forecast of El Niño, our algorithm instead employs the precursors in the dynamical strength of the teleconnections in the climate network to predict the onset of the warming.

Our results suggest that for enabling local perturbations of the environment to instigate an El Niño event, the network needs to be in a "cooperative" state that can be characterized, to a certain extent, by sufficiently large link strengths in the considered climate network. The cooperativity sets in *well before* the spring barrier and thus allows for an early forecasting of ENSO. This situation might be related to the mechanism suggested in [39] for optimal SST growth, which is essentially the emergence of a certain spatial SST pattern, resembling our finding of a cooperated fluctuation.

To study the robustness of the forecasting algorithm with respect to the underlying network structure, we varied the size of the El Niño basin by (i) eliminating the grid point below the equator in the middle of the Pacific (thus identifying the El Niño-basin with the NINO1,2,3 and 3.4 regions); and (ii) equating the El Niño-basin with the NINO1,2,3 and 4 regions (thus raising the number of the grid points in the basin to 17). In addition, we diluted the network by connecting only 20% of the surrounding nodes with the El Niño-basin. We found that with all these modifications the performance of the forecasting algorithm was only slightly reduced (see Figs. 7 - 11 in the supplemental material). We also tested the performance of the algorithm when the outer grid points are not in the Pacific region but in Europe. The performance of this network was considerably weaker and comparable with the incumbent 12-months forecasts (see Fig. 3 and Fig. 12 in the supplemental material). Finally, we tested if our algorithm can also forecast high levels (above the standard deviation) of the negative standard southern oscillation index (SOI), which is strongly correlated to the NINO3.4 index (see, e.g., [40]). In contrast to the NINO3.4 index, the SOI uses atmospheric data, the pressure difference between Tahiti and Darwin. In the supplemental material we show (Fig.13) that our algorithm is also able to forecast high levels of the negative SOI, with a hit rate close to 0.67 at a false alarm rate close to 0.17.

In summary, we propose a climate-network approach to forecast El Niño-episodes about one year ahead. Our approach is based on the dynamic fluctuations of the teleconnections (links in the network) between grid points in the El Niño-basin and the rest of the Pacific. The strengths of the links are obtained from the cross-correlations between the observed sea-surface-level air temperatures in the grid points. We have shown explicitly that our method outperforms existing methods in predicting El Niño-events at least 6-12 months in advance. In contrast to the algorithms using model data, our method is exclusively based on instrumental accounts which are easily accessible. Thus the results of this study can be straightforwardly reproduced.

We did not aim to forecast La Nina events, where the NINO3.4 index is below -0.5°C for more than 5 months. In a trivial forecast, one predicts that an El Nino event will be followed by a La Nina event in the next year. This simple forecast has, in the considered time window between 1950 and 2012, a hit rate of 0.73 and a false alarm rate of 0.17. An even better forecast of La Nina events using the climate network requires an additional precursor to be found and is beyond the scope of this Letter.

Altogether, our findings indicate that El Niño is a cooperative phenomenon where the teleconnections between the El Niño-basin and the rest of the Pacific tend to build up in the calendar year before an event. For characterizing the teleconnections we have used a univariate model where only one climate variable (atmospheric temperature) has been used. In the future, we plan to extend this model to the multivariate case where other climate variables (pressure, wind speed etc) are taken into account as well.

Finally, we would like to note that our algorithm (see Fig. 2b) did correctly predict the absence of an El Niño-event in 2012. This forecast was made in 2011 already, whereas conventional approaches kept on predicting the warming occurrence far into the year 2012 [41].

*Acknowledgement:* We acknowledge financial support from the Deutsche Forschungsgemeinschaft.